# Deposition of nanosized amino acid functionalized bismuth oxido clusters on gold surfaces

*Annika Morgenstern[1], Rico Thomas[2], Apoorva Sharma[1], Marcus Weber[2], Oleksandr Selyshchev[1], Ilya Milekhin[1], Doreen Dentel[4], Sibylle Gemming[1,3], Christoph Tegenkamp[4], Dietrich R.T. Zahn,[1,3], Michael Mehring[2,3], Georgeta Salvan[1,3]*

[1] Semiconductor Physics, Chemnitz University of Technology, 09107 Chemnitz, Germany

[2] Coordination Chemistry, Chemnitz University of Technology, 09107 Chemnitz, Germany

[3] Center of Materials, Architectures and Integration of Nanomembranes, Chemnitz University of Technology, 09126 Chemnitz, Germany

[4] Analysis of Solid Surfaces, Chemnitz University of Technology, 09107 Chemnitz, Germany

**Abstract**

Bismuth compounds are of growing interest with regard to potential applications in catalysis, medicine and electronics, for which their environmentally benign nature is one of the key factors. The most common starting material is bismuth nitrate, which easily hydrolyses to give a large number of condensation products. The so-called bismuth subnitrates are composed of bismuth oxido clusters of varying composition and nuclearity. One reason that hampers the further development of bismuth oxido-based materials, however is the low solubility of the subnitrates, which makes targeted immobilisation on substrates challenging. We present an approach towards solubilisation of bismuth oxido clusters by introducing an amino carboxylate as functional group and a study of the growth mode of these atom-precise nanoclusters on gold surfaces. For this purpose the bismuth oxido cluster [$Bi_{38}O_{45}(NO_3)_{20}(dmso)_{28}$]($NO_3$)$_4$·4dmso (dmso = dimethyl sulfoxide) was reacted with the sodium salt of *tert*-butyloxycabonyl (Boc)-protected phenylalanine (Phe) to give the soluble and chiral nanocluster [$Bi_{38}O_{45}$(Boc–Phe)$_{24}$(dmso)$_9$]. The hydrodynamic diameter of the cluster was estimated with (1.4 – 1.6) nm (in $CH_3CN$) and (2.2 nm – 2.9) nm (in ethanol) based on dynamic light scattering (DLS). The full exchange of the nitrates by the amino carboxylates was proven by nuclear magnetic resonance (NMR) and Fourier-transform infrared spectroscopy (FTIR) as well as elemental analysis (EA) and X-ray photoemission spectroscopy (XPS). The solubility of the bismuth oxido cluster in a protic as well as an aprotic polar organic solvent and the growth mode of the clusters on Au upon spin, dip, and drop coating on gold surfaces were studied. Successful deposition of bismuth oxido cluster was proven by powder X-ray diffraction (PXRD), FTIR, and XPS while the microstructure of the resulting films was investigated as a function of the deposition method and the solvent used by scanning electron microscopy (SEM), atomic force microscopy (AFM), and optical microscopy. In all cases, the bismuth oxido clusters form crystalline agglomerations and their size, height, and distribution can be controlled by the choice of the solvent and of the deposition method.

**Introduction**

The environmentally benign nature of bismuth compounds offers access to novel sustainable materials and led to developments *e.g.* in catalysis, medicine, radiopaque additives, and electronics.[1-12] Among the starting materials used for the synthesis of solid state materials, bismuth(III) nitrate pentahydrate is one of the most common and readily available compounds. However, it shows poor solubility in all kinds of solvents and easily hydrolyses to give various condensation products of different stoichiometry and nuclearity.[13-15] We studied these so-called bismuth subnitrates recently in order to elucidate the molecular mechanisms of the condensation that lead to the formation of bismuth oxido nitrates with up to 38 bismuth atoms, *e.g.* in the nanoscaled [$Bi_{38}O_{45}(NO_3)_{24}$], using a combined *in-situ* small-angle X-ray scattering (SAXS) and pair distribution function (PDF) analysis approach as well as electrospray ionization mass spectrometry (ESI-MS).[16-18] The $Bi_{38}O_{45}$ core seems to be the most stable one among the bismuth oxido clusters and various molecules of this type were reported recently including sulfonates, salicylates, nitrates, and methacrylates.[13-14, 19-24] However, most of them suffer from either low hydrolytic stability at the ligand periphery, low solubility, low biocompatibility of their ligands, or accessibility on a large scale.[18, 25] Previously, we demonstrated that the ligand substitution starting from [$Bi_{22}O_{26}(OSiMe_2^tBu)_{14}$] using *tert*-butyloxycabonyl (Boc)-protected phenylalanine (Phe) (Boc–Phe–OH) gave [$Bi_{22}O_{26}$(Boc–Phe–O)$_{14}$], which quickly hydrolyses to result in the formation of [$Bi_{38}O_{45}$(Boc–Phe–O)$_{22}$(OH)$_2$].[26] Furthermore, we showed that the substitution of nitrate anions in [$Bi_{38}O_{45}(NO_3)_{20}$(dmso)$_{28}$](NO$_3$)$_4$·4dmso is possible by the reaction with sodium methacrylate to give [Bi$_{38}$O$_{45}$(OMc)$_{24}$] (OMc = Methacrylate) as proven by *in-situ* studies.[16, 21] The combined hydrolysis and substitution reaction starting from bismuth nitrate and sodium carboxylates usually tends to give a [Bi$_{38}$O$_{45}$] cluster core surrounded by a mixed ligand shell of nitrates and carboxylates.[27] Here we report the straightforward synthesis of the chiral and highly soluble bismuth oxido cluster [Bi$_{38}$O$_{45}$(Boc–Phe–O)$_{24}$(dmso)$_9$] by a simple substitution of the nitrate anions in [Bi$_{38}$O$_{45}$(NO$_3$)$_{20}$(dmso)$_{28}$](NO$_3$)$_4$·4dmso by the reaction with Boc–Phe–ONa with a high yield. The solubility of this compound and its functional chiral ligand shell makes it attractive for the deposition on diverse substrates. This step is essential for developments in the field of thin films and their applications based on metal oxido clusters, as demonstrated in the rich field of polyoxometalates (POMs) with potential applications ranging from medicine and catalysis[28-29] over energy conversion[30] to electronic-[31-32] and electrochromic devices.[28] Reports on deposition methods for neutral metal oxido clusters are so far restricted to rare examples for metal oxido clusters with titanium, tin, hafnium, zirconium, and the lanthanides.[33-37]

The deposition of bismuth oxido nanoclusters on surfaces is almost unexplored. We spray-coated [Bi$_{38}$O$_{45}$(OMc)$_{24}$(dmso)$_9$] on glass surfaces for thin film formation of bismuth(III) oxide after subsequent calcination.[38] Recently, Wu *et al.* embedded hexanuclear bismuth oxido clusters of the type [Bi$_6$O$_4$(OH)$_4$(NO$_3$)$_6$] in a carbon composite material showing a good capacity as anode materials in lithium-ion batteries,[33] while Zhou *et al.* reported on thin films of similar hexanuclear bismuth oxido/hydroxide nitrates on a glass substrate, covered by a thin aluminium layer. The bismuth oxido clusters showed potential for the use in rewritable resistive memory device.[39] To the best of our knowledge, the deposition of larger bismuth oxido clusters on a metallic substrate has not yet been reported. Nevertheless, the deposition is a requirement for applications in, *e.g.* in electronic devices, or sensors as reported for other clusters on gold surfaces.[40-42]

We report here on the growth mode of the Boc-protected amino acid functionalized chiral bismuth oxido nanocluster [$Bi_{38}O_{45}$(Boc–Phe–O)$_{24}$(dmso)$_9$] on Au-coated silicon substrates using spin, drop, and dip coating methods. The film morphology was analysed using optical microscopy, scanning electron microscopy (SEM) and atomic force microscopy (AFM). The samples were characterised using X-Ray diffraction (XRD), FTIR-, and XP-spectroscopies to verify the structural and chemical integrity of the nanoclusters after deposition from solution.

 Experimental section

**Materials**

*N*-(*tert*-butoxycarbonyl)-L-phenylalanine (Boc–Phe–OH) was purchased from Sigma Aldrich and used without further purification. $Bi(NO_3)_3·5H_2O$ was purchased from Alfa Aesar and used without further purification. [$Bi_{38}O_{45}(NO_3)_{20}$(dmso)$_{28}$]($NO_3$)$_4$·4dmso (**A**) was synthesized according to a literature procedure.[13] $Na_2CO_3$ (99.5 %) from Sigma Aldrich was used without further purification. Acetonitrile purchased from Merck and ethanol (Alfa Aeser) with 99 % purity were used without further purification.

**Synthesis of the sodium salt of Boc-Phe-OH (1)**

*N*-(*tert*-butoxycarbonyl)-L-phenylalanine (1.500 g, 5.65 mmol) was dispersed in deionized water (15 ml) under ambient conditions. $Na_2CO_3$ (300 mg, 2.83 mmol) was added slowly and gas evolution occurred immediately when the suspension was ultrasonicated for a few seconds ($f$ = 35 kHz). The solution was heated to 80 °C for 1 h. After filtration and cooling to room temperature, a colorless solid of Phe–Boc–ONa (**1**, 1.394 g, 4.85 mmol, 85 %) was obtained after slow evaporation of the solvent.

$^1$H NMR (ppm, 500.30 MHz, dmso-$d_6$, 298 K) $\delta$ = 7.18 (m, 5 H), 5.85 (d, 1 H) 3.81 (q, 1 H), 3.05 (dd, 1 H), 2.88 (dd, 1 H), 2.50 (q, dmso-$d_6$) 1.32 (s, 9 H). $^{13}$C NMR (ppm, 125.80 MHz, dmso-$d_6$, 298 K) 173.9, 154.7, 139.4, 129.7, 127.8, 125.7, 77.4, 56.4, 37.7, 28.4. CHNS (%,

expt. and calcd.) for $C_{14}H_{18}O_4NNa$ (M = 287.32 g·mol$^{-1}$): C, 58.68 (58.52); H, 6.34 (6.33); N, 4.82 (4.88). IR (cm$^{-1}$) 3330 m, 3063 w, 3031 w, 2975 m, 2930 w, 1680 m, 1583 s, 1497 m, 1452 w, 1389 s, 1365 s, 1250 m, 1163 s, 1048 m, 1024 m, 852 w, 754 m, 698 s, 561 m, 464 m.

**Synthesis of [Bi$_{38}$O$_{45}$(Boc-Phe-O)$_{24}$(dmso)$_9$] (2)**

The bismuth oxido cluster [Bi$_{38}$O$_{45}$(NO$_3$)$_{20}$(dmso)$_{28}$](NO$_3$)$_4$·4dmso (**A**, 1.000 g, 0.08 mmol) was dispersed in dmso (40 ml) and heated to 80 °C for 1 h to give a colorless solution. Phe-Boc-ONa (**1**, 837 mg, 2.916 mmol) was added and the colorless solution was stirred at 80 °C for 4 h. The hot solution was filtrated and allowed to cool to room temperature. A colorless solid of **2** was obtained after slow evaporation of the solvent for a few days. The solid was washed with 15 ml of deionized water and then dried under ambient conditions for two days. Compound **2** was obtained as a colorless solid (1.252 g, 0.08 mmol, 99 % based on bismuth in **A**).

$^1$H NMR (ppm, 500.30 MHz, dmso-d$_6$, 298 K) $\delta$ = 7.15 (m, 5 H), 6.12 (d, 1 H) 3.91 (q, 1 H), 3.07 (dd, 1 H), 2.84 (dd, 1 H), 2.54 (s, dmso$_{coord.}$) 2.50 (q, dmso-d$_6$) 1.28 (s, 9 H). $^{13}$C NMR (ppm, 125.80 MHz, dmso-d$_6$, 298 K) 173.9, 155.0, 139.2, 129.6, 127.9, 125.9, 77.6, 56.4, 40.4 (dmso$_{coord.}$), 37.7, 28.4. CHNS (%, expt. and calcd.) for Bi$_{38}$O$_{150}$C$_{354}$H$_{486}$S$_9$N$_{24}$ (M = 15708,27 g·mol$^{-1}$): C, 26.60 (27.06); H, 3.19 (3.13); N, 2.82 (2.14); S, 1.81 (1.84). IR (cm$^{-1}$) 3330 m, 3060 w, 3029 w, 2975 m, 2928 w, 1688 m, 1554 m, 1465 s, 1452 w, 1384 s, 1363 s, 1246 m, 1162 s, 1047 m, 1018 m, 948 w, 854 w, 753 m, 699 s, 528 s.

**Film preparation**

The bismuth oxido cluster **2** was dissolved in ethanol as protic, polar solvent and for comparison in acetonitrile as aprotic, polar solvent. A concentration of 20 g·l$^{-1}$ was applied for all samples. As substrates 10 mm x 10 mm pieces of silicon wafers with a 150 nm silicon oxide layer, a 20 nm sputtered titanium layer as an adhesion layer, and 100 nm gold (Au) were used.

Drop coating was performed by covering the substrate with 20 µl of solution and evaporation of the solvent under ambient conditions.

Spin coating was performed using a Laurell WS-650MZ-23NPPB spin coater with a 4.5 mm vacuum chuck and a pressure of 1 mbar. The coater was operated with a rotation time of 30 s at a rotation speed of 2000 rpm. 20 µl of the solution was dropped onto the substrate before the rotation was started.

The dip coating experiments were performed using a SOLGELWAY ACEdip 2.0 dip coater. The samples were dipped into 3 ml of the solution at a constant speed of 5 mm·s$^{-1}$ for 100 s. Afterwards the solution was removed with a constant speed of 2 mm·s$^{-1}$ and dried under ambient conditions.

The parameters used for the ethanol and acetonitrile samples are shown in Table 1, were the sample same is composed of the **2** compound, the deposition technique and the solvent, respectively.

Table 1. Deposition parameters for the samples discussed in this work

| sample name | deposition method | solvent | rotation speed (rpm) | dip/ withdrawing speed (mm·s$^{-1}$) | amount of solution (µl) |
|---|---|---|---|---|---|
| **2**-SpA | spin coating | acetonitrile | 2000 | - | 20 |
| **2**-SpE | spin coating | ethanol | 2000 | - | 20 |
| **2**-DiA | dip coating | acetonitrile | - | 5 / 2 | - |
| **2**-DiE | dip coating | ethanol | - | 5 / 2 | - |
| **2**-DrA | drop coating | acetonitrile | - | - | 20 |
| **2**-DrE | drop coating | ethanol | - | - | 20 |
| **2**-DrA-2 | drop coating | acetonitrile | - | - | 50 |
| **2**-DrE-2 | drop coating | ethanol | - | - | 50 |

**Characterization methods**

Optical microscopy imaging was performed with a Nikon IC inspection Microscope ECLIPSE L200 with 1x as well as 50x objectives. Atomic force microscopy (AFM) measurements were performed with an atomic force microscope (NanoWorld Arrow$^{TM}$NCPt AFM) operating in intermediate topography mode using a *PtIr$_5$* cantilevers with resonant frequencies in the range of (240 – 380) kHz. The analysis of the surface coverage was performed either using Image J[43] (for the optical microscopy images) or Gwyddion software[44] (for the AFM images). Fourier transformed infrared (FTIR) spectroscopy was performed in the wavenumber range of (450 – 5000) cm$^{-1}$ using a VERTEX 80v FTIR spectrometer with an attenuated total reflectance (ATR) unit. X-ray photoelectron spectroscopy (XPS) analysis was performed with an ESCALAB 250Xi photoelectron spectrometer from Thermo Scientific$^{TM}$ in an ultra-high vacuum (UHV) chamber using a monochromatic Al-K$\alpha$ (1486.68 eV) X-ray source and a beam diameter of 300 $\mu$m. The binding energies of all spectra were referenced to the binding energy of C1s (248.8 eV). UV–vis spectra were measured in the diffuse reflection geometry at ambient temperature with an Agilent Technologies Cary 60 UV–vis spectrometer in a range from 200 nm to 800 nm using an optical probe. The data were collected with the *Cary WinUV* Software (version 5.0.0.1005). Powder X-ray diffractograms were measured at ambient temperature with an STOE *Stadi P* diffractometer (Darmstadt, Germany) using Ge(111)-monochromatized Cu-K$_\alpha$ radiation (1.54056 nm, 40 kV, 40 mA). The full width at half maximum (FWHM) is corrected for instrumental broadening using a LaB$_6$ standard (*SRM 660*) purchased from the National Institute of Standards and Technology (NIST). The value $\beta$ was corrected using $\beta^2 = \beta_{measured}^2 - \beta_{instrument}^2$ where $\beta_{measured}^2$ and $\beta_{instrument}^2$ are the FWHMs of measured and standard profiles, respectively. Particle size distribution (PSD) based on dynamic light scattering (DLS) was determined using a Zetasizer Nano ZS (Malvern Instruments) allowing the characterization of suspension in the size range of 0.6 nm to 6 μm. A red laser

(633 nm, 4 mW) was used as light source and the analyses were performed at an angle of 173 ° (backscatter NIBS default). The compounds were dissolved in appropriate solvents (20 g·L$^{-1}$), filtered, filled into glass cuvettes (DTS0012), and measured at 20 °C. Calculation of the PSD was carried out according to the "Mie theory" assuming the presence of spherical particles. $^1$H and $^{13}$C{$^1$H} NMR spectra were recorded at room temperature in dmso-d$_6$ (dried over 4 Å molecular sieve) with an *Avance III 500* spectrometer (BRUKER) at 500.30 Hz and 125.80 MHz, respectively, and were referenced internally to the deuterated solvent relative to Si(CH$_3$)$_4$ ($\delta$ = 0.00 ppm). Scanning electron microscopy was performed using a NanoNovaSEM200 (ThermoFisher) device with an electron beam energy of 5 keV and different magnifications.

**Results and Discussion**

**Cluster synthesis and characterisation**

Starting from [Bi$_{38}$O$_{45}$(NO$_3$)$_{20}$(dmso)$_{28}$](NO$_3$)$_4$·4dmso (**A**) and the sodium salt of Boc–L–Phe–OH (**1**) in dmso, [Bi$_{38}$O$_{45}$(Boc–Phe–O)$_{24}$(dmso)$_9$] (**2**) was obtained after a few days of crystallization under ambient conditions as colorless solid with yields > 90 %. Elemental analysis confirms the stoichiometry of compound **2**, complete ligand substitution was proven using and NMR in combination with IR spectroscopy. The $^1$H-NMR spectrum reveals all expected signals for the Boc–Phe–O$^-$ ligand and is indicative for dmso coordination to the bismuth oxido cluster in solution (dmso-d$_6$) showing a characteristic signal at $\delta$ = 2.54 ppm. The IR spectrum shows all expected vibrations of the Boc-protected amino acid- based ligands (Figure S1). Compared to the sodium salt **1**, the bismuth oxido cluster **2** shows a characteristic broad Bi–O vibration at 580 cm$^{-1}$.[45] The absence of vibrations at approximately 1740 cm$^{-1}$ ($\nu$(N=O)$_{free}$) and 1270 cm$^{-1}$ ($\nu_{sym.}$(NO$_2$)) demonstrates the complete exchange of the nitrate ligands.[46-47] The symmetric valence S=O vibration at 948 cm$^{-1}$ is indicative for dmso in the resulting compound, which is in line with the NMR results.[48] The presence of Boc–Phe–OH is confirmed by the $\delta_{C-H}$ vibration of the *tert*-butyl group at 1386 cm$^{-1}$ and 1363 cm$^{-1}$, the $\delta_{C=C-H}$ vibration for a monosubstituted aromatic ring at 699 cm$^{-1}$, 753 cm$^{-1}$, and 1018 cm$^{-1}$ and the $v_{C-N}$ as well as the $v_{C-O}$ at 1162 cm$^{-1}$ and 1047 cm$^{-1}$, respectively.[45, 49-50] All these vibrations are only slightly shifted compared to Boc–Phe–OH and the sodium salt **1**, however, the symmetric and asymmetric $v_{C=O}$ at 1465 cm$^{-1}$ and 1554 cm$^{-1}$ show significant shifts. Especially the asymmetric vibration in **2** is shifted to lower wavenumbers by approximately 30 cm$^{-1}$ compared to **1** and by 90 cm$^{-1}$ compared to Boc–Phe–OH, which indicates different coordination to the cluster core.

The preservation of the [Bi$_{38}$O$_{45}$] cluster core upon the ligand exchange reaction is confirmed using PXRD (Figure S2) which shows a typical diffraction pattern for bismuth oxido nanoclusters showing one intensive diffraction peak and a much weaker one for $2\theta < 10\,°$.[18, 27] Compared to the reflections of [Bi$_{38}$O$_{45}$(NO$_3$)$_{20}$(dmso)$_{28}$](NO$_3$)$_4$·4dmso ($2\theta = 5.23\,°$ and $2\theta = 6.24\,°$) the main diffraction peak of **2** is shifted to a lower angle at $2\theta = 4.47\,°$, indicating an increase of the cluster size of **2** due to the larger ligand shell compared with the nitrate cluster. The previously reported [Bi$_{38}$O$_{45}$(Boc–Phe–OH)$_{22}$(OH)$_2$] shows similar diffraction peaks with $2\theta = 4.51\,°$ and $2\theta = 5.19\,°$.[26]

The main reflection corresponds to the interlayer distance assuming a closed packed nanocluster arrangement.[20, 51] Thus, an interlayer distance for **2** of 1.975 nm ($2\theta = 4.47\,°$) was calculated using Bragg's equation. Compared to the precursor [Bi$_{38}$O$_{45}$(NO$_3$)$_{20}$(dmso)$_{28}$](NO$_3$)$_4$·4dmso (**A**) with its main reflection at $2\theta = 5.23\,°$ ($d = 1.688$ nm) the resulting interlayer distance increases according to the molecule size. In dmso solution, the mean hydrodynamic diameter was observed in the range of $d_h = (2.2 – 2.9)$ nm and of $d_h = (1.2 – 1.5)$ nm for compounds **2** and **A**, respectively. The hydrodynamic diameter was determined from fitted particle size distribution (PSD) curve based on dynamic light scattering (DLS) the exemplarily results for each solvent are shown in Figure S3. The DLS analysis for the nitrate-substituted bismuth oxido cluster **A** is restricted to the usage of dmso as solvent as a result of its low solubility, whereas cluster **2** is soluble in various organic solvents. Using ethanol as solvent, the resulting hydrodynamic diameter of **2** is in the range of $d_h = (2.2 – 2.9)$ nm and of $d_h = (1.4 – 1.6)$ nm, when acetonitrile was used. In the former case, we assume that the bismuth oxido cluster is solvated by the ethanol molecules providing an increased periphery. In acetonitrile, partial dissociation of coordinating ligands and solvent, as well as partial hydrolysis and partial deprotection of the BOC-group are assumed to give smaller cluster molecules and an asymmetrical molecule shape, which results in a reduced hydrodynamic diameter. Noteworthy, the solubility of **2** is twice as high in ethanol ($\beta = 65$ g·L$^{-1}$) as in acetonitrile ($\beta = 30$ g·L$^{-1}$). However, the cluster core of **2** is stable in both solvents, which make it suitable for the investigation of solution-based deposition methods. The influence of solubility as well as of the stabilization and partial dissociation of the cluster **2** in solution onto the growth mode of the films on Au is discussed in the following.

In order to test the chemical and structural integrity of compound **2** upon dissolving in various solvents followed by deposition (dip coating, spin coating and drop coating) onto the Au-coated substrates, XRD, ATR-FTIR, and XPS measurements were performed. In the following, the observations made upon dip coating are exemplarily discussed.

The grazing incidence XRD results of the samples deposited by dip coating are shown in Figure 1, and compared to the reference PXRD pattern for **2** measured in the powder diffraction mode. A small shift of the reflections might be expected considering a slightly different crystal packing in the powder and in the films. The XRD patterns over the full range from $2\theta = 3\,°– 85\,°$ are dominated by Au(111), Au(222) and Si(400) reflections. For **2-DiE** and **2-DiA** only one strong reflection at $2\theta = 4.38\,°$ and $2\theta = 4.35\,°$, respectively, is observed, which corresponds to the main reflection peak of **2** at $2\theta = 4.47\,°$. Thus, we conclude that the deposition of compound **2** was successful.

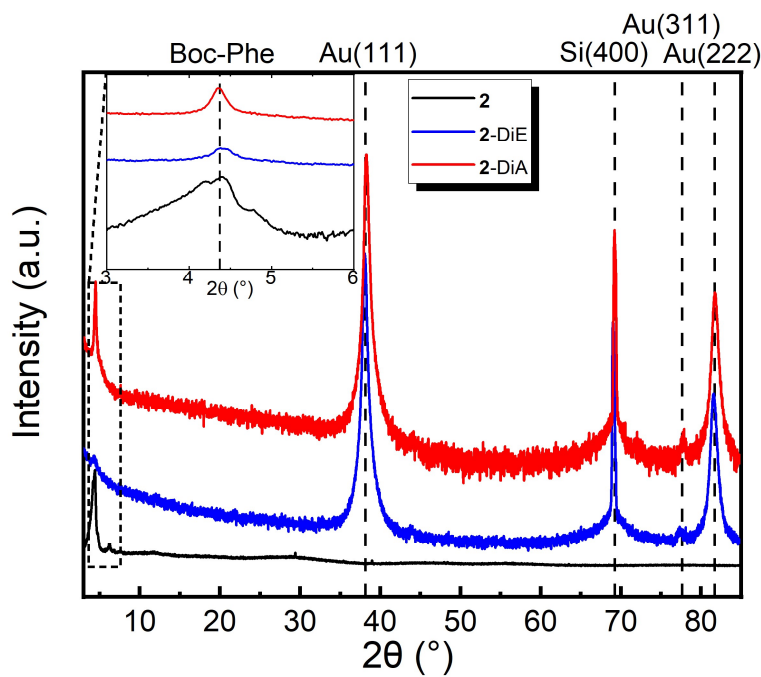

Figure 1. XRD patterns of dip coated thin films **2-DiE** and **2-DiA** compared to the PXRD of **2**. The intensity is shown on a logarithmic scale. The inset shows a zoom in the range where the characteristic signature of crystals of **2** is expected.

XPS measurements were carried out to analyse the chemical composition of the films. The survey spectrum (cf. Figure 2a) of a representative sample deposited by dip coating from acetonitrile solution (**2-DiA**) reveals the typical signals for bismuth (Bi 5d, Bi $5p_{3/2}$, Bi $5p_{1/2}$, Bi 4f, Bi $4d_{5/2}$, Bi $4d_{3/2}$, Bi $4p_{3/2}$, Bi $4p_{1/2}$), oxygen (O 1s) as well as carbon (C 1s).[47, 52-53] A peak at 399.7 eV arising from nitrogen N 1s in the amide groups[54] was observed in the high resolution spectrum (Figure 2b). By contrast the N 1s peak for the starting material $[Bi_{38}O_{45}(NO_3)_{20}(dmso)_{28}](NO_3)_4 \cdot 4dmso$ (**A**) is observed at 405.8 eV (cf. Figure 2b), which indicates the complete ligand exchange from nitrate to the Boc-protected amino acid carboxylate.[47] In the high resolution spectrum for O 1s, the main peak at 531.4 eV is associated to the C−O bond; the satellite peak at a higher binding energy of 532.9 eV corresponds to the

C=O bond (cf. Figure 2c).[55] The peak at the lower binding energy of 530.1 eV stems from Bi–O bonds in the cluster core[56] (Figure 2c), which proves the existence of the bismuth oxido core after the deposition on the Au-coated substrate. In Figure 2d the high-resolution spectrum for C 1s is shown. The main peak corresponds to the $sp^3$ hybridized carbon of the Boc-protecting group as well as the $CH_3$ group in Phe and dmso.

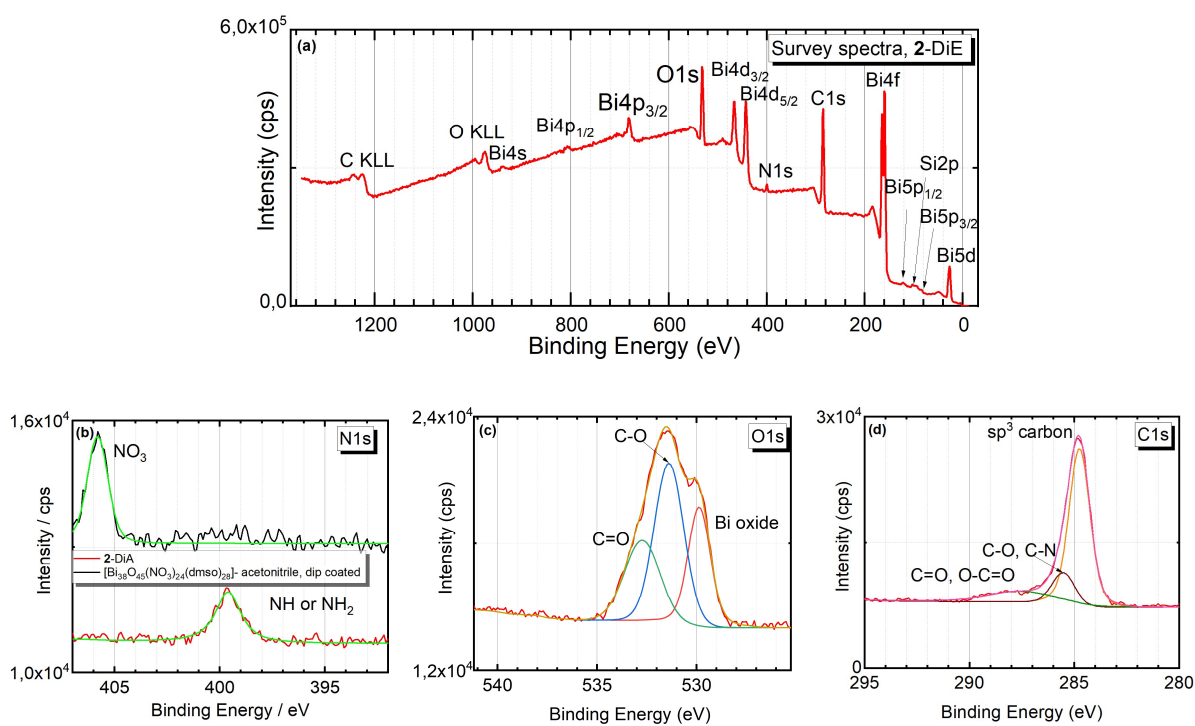

Figure 2. (a) Survey X-ray photoemission spectra of **2**-DiA, high resolution spectra for (b) N 1s comparison between **2**-DiA and **A**; a clear indication of successful ligand exchange is provided by the clear shift to lower binding energies of (c) O 1s and (d) C 1s core levels.

The FTIR spectra of **2**-DiA and **2**-DiE along with that of **2** are shown exemplarily in Figure 3 and the assigned vibrations are listed in Table S1**Fehler! Verweisquelle konnte nicht gefunden werden.**. FTIR data for spin and drop coated samples are shown in Figure S4.

For all samples, the most intensive vibrational bands were observed in the range from 570 cm$^{-1}$ – 1700 cm$^{-1}$. The broad band at approximately 582 cm$^{-1}$ is assigned to the stretching vibration of Bi–O,[45] which is slightly broadened and shifted to higher wavenumbers in the films than in the powder of **2**. In contrast, the bands assigned to the Boc-protected amino acid do not show significant changes. The bands observed at 700 cm$^{-1}$, 753 cm$^{-1}$ and 1020 cm$^{-1}$ correspond to deformation vibrations of the mono-substituted aromatic system from Phe unit.[45, 49] The intense vibrational modes at 1363 cm$^{-1}$ and 1386 cm$^{-1}$ belong to the symmetric bending mode of C−H (methyl) groups.[50] The bands appearing at 1495 cm$^{-1}$ and 1554 cm$^{-1}$ correspond to the symmetric C=O stretching of carboxylate and at 1688 cm$^{-1}$ to the C=O stretching of amide. The aromatic C−H stretching vibrations above 3000 cm$^{-1}$ and the aliphatic C−H stretching vibration below 3000 cm$^{-1}$ can be assigned to the Boc–Phe–O$^-$ ligands.[50] The characteristic symmetric S=O vibration band of compound **2** located at approximately 948 cm$^{-1}$ is not present in the films.[48] We conclude that dmso as solvate of compound **2** is removed upon dissolving in ethanol or acetonitrile, respectively, due to the very low concentration of dmso in the resulting solutions. This behaviour was recently reported for [Bi$_{38}$O$_{45}$(OMc)$_{24}$(dmso)$_9$]·2dmso·7H$_2$O, for which the prior solvate dmso was displaced by appropriate alcohol solvate molecules according to the solvent used for recrystallization (EtOH or $^i$PrOH).[57]

In conclusion, **2** was solubilized in the considered solvents and deposited from solution while retaining the main composition and cluster structure except for the substitution of coordinating solvent dmso molecules, whose vibrational signature is not observed in the deposited films. In the following we discuss the macroscopic growth mode of the clusters on gold surfaces using optical - and scanning electron microscopy.

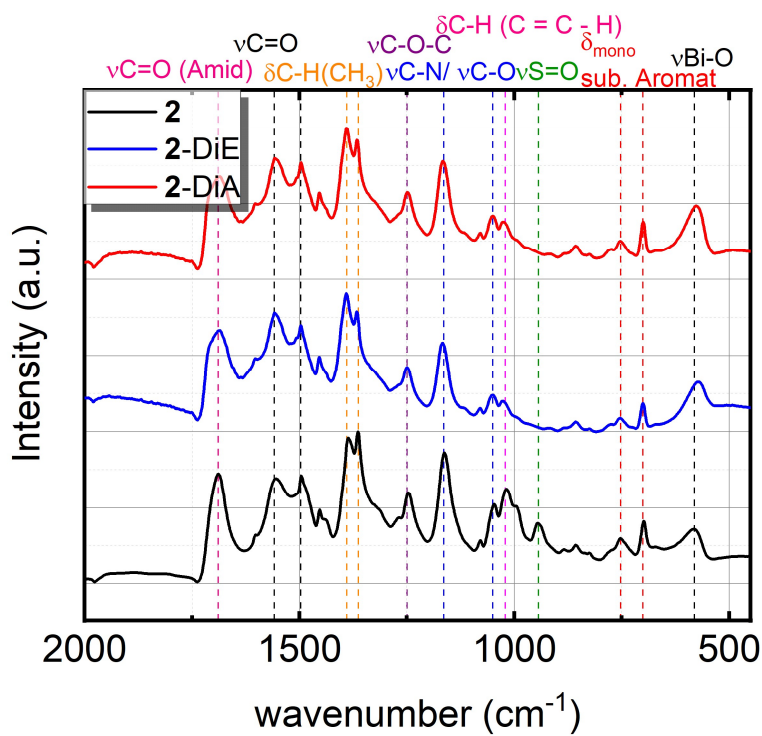

Figure 3. FTIR spectra of thin films of **2** deposited by dip coating from acetonitrile (**2**-DiA) or ethanol (**2**-DiE) solution and comparison to pure **2**.

**Drop coated films**

The contact angle of the droplets of the investigated solutions on the Au surface varied from 2 ° to 8 °, which indicates wetting between the Au substrate and all solvents used. According to ref.[58] three main competing flow patterns can occur in droplets of solutions containing colloidal particles when deposited onto a substrate. On one hand, a fast evaporation at the wetting border of the droplet to the unwetted substrate induces a radial flow from the middle of the droplet towards the border, leading to a ring-like agglomeration of the nanoparticles. On the other hand, the presence of a dominant Marangoni recirculating flow would lead to a strong agglomeration of nanoparticles in the middle of the droplet. Third, the transport of particles towards the substrate driven by Derjaguin-Landau-Verwey-Overbeek (DLVO) interactions would lead to a strong agglomeration of nanoparticles in the middle of the droplet or to a highly homogeneous coverage of the substrate, respectively. Depending on the thermophysical properties of the substrate, on the concentration, and on the viscosity as well as on the pH-value of the solution, various structural properties of the films formed by the nanoclusters remaining on the substrate after the solvent evaporation can be achieved. Despite the fact that the phase diagram in ref.[58] was proposed for colloidal nanoparticles with dimensions which are two orders of magnitude larger than those of compound **2**, some similarities between the agglomeration patterns observed in this study and those reported in [58] can be observed.

For a better understanding of the difference in growth behaviour on the edges and the middle of the droplet, SEM images are shown in Figure 4 for **2-DrE** films and in Figure 5 for **2-DrA** for films deposited from 20 µl solution.

For low viscosity solvents ($\eta$) with a high evaporation rate ($\Gamma$) like ethanol ($\eta(EtOH) = 1.2$ mPa; $\Gamma(EtOH) = (0.12 \pm 6.00 \cdot 10^{-3})$ µl·s$^{-1}$·cm$^{-2}$ and acetonitrile ($\eta(CH_3CN) = 0.4$ mPa; $\Gamma(CH_3CN) = (0.08 \pm 4.10 \cdot 10^{-3})$ µl·s$^{-1}$·cm$^{-2}$)[59] the turbulences influence the growth behaviour significantly, but the trade-off to pay is the inhomogeneity of the film thickness, which was determined by profilometry. The line profiles obtained of the samples drop coated from ethanol (**2-DrE**) and acetonitrile (**2-DrA**) solution are shown in Figure S7. The maximum height of agglomerations of **2** is observed at the borders of the dried droplet in the case of acetonitrile, while compound **2** is more evenly distributed in the case of the deposition from the ethanol solution. The tendency for agglomeration can be understood as a consequence of the stronger intermolecular interaction between the bismuth oxido clusters in solution compared to rather weak interactions between **2** and the gold surface.[60]

On the edges of the initial droplets (cf. Figure 4b and Figure 5b) rough agglomerations can be observed. Judging from the form and size of the bismuth oxido cluster agglomerations it can be inferred that in the middle of the droplets (cf. Figure 4c and Figure 5c) the dominant growth mode is the coffee stain effect[58] for both solvents discussed here. Due to the low evaporation rate, agglomerations in the middle of the droplet are observed as well (cf. Figure 4c and Figure 5c).

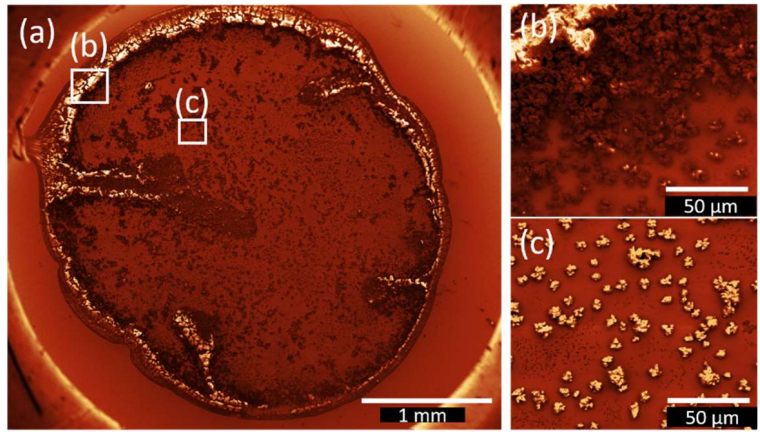

Figure 4. SEM image of **2-DrE** film after the evaporation of the solvent and treated with the Gwyddion.net image filter from Gwyddion software, (a) the film shows characteristic features of the coffee stain effect in combination with Ostwald ripening, (b) indicates rough film edges and large agglomerations of bismuth oxido clusters at the drop edge, and (c) displays the middle part of the film with agglomerations of different size next to each other and the trend to more pronounced depletion zones around the larger agglomerations, which is an indicator for Ostwald ripening.

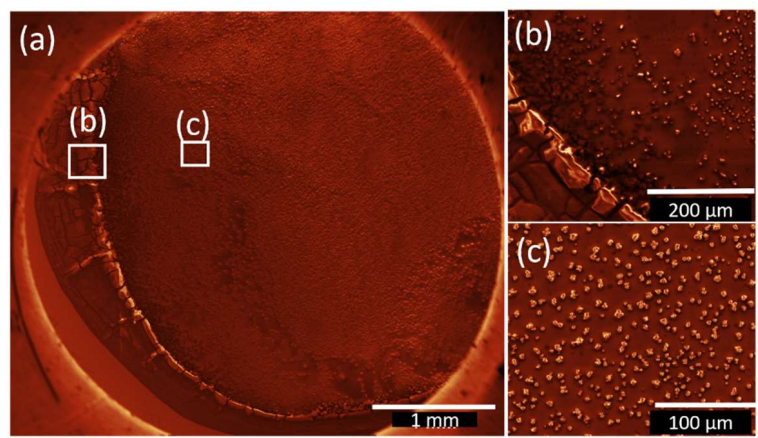

Figure 5. SEM image of **2-DrA** film after the evaporation of the solvent and treated with the Gwyddion.net image filter from Gwyddion software, (a) full drop shows the coffee stain effect, (b) indicates rough film edges as well as large cluster agglomerations at the drop edge, and (c) displays the middle part of the droplet with homogeneously distributed agglomerations of similar size.

Furthermore, experiments with higher amount of solution (50 µL), covering the whole substrates were made to determine the difference in growth mode shown in Figure 6**Fehler! Verweisquelle konnte nicht gefunden werden.** and Figure 7**Fehler! Verweisquelle konnte nicht gefunden werden.** for ethanol (**2**-DrE-2) and acetonitrile (**2**-DrA-2), respectively. As can be seen from the microscopy image in Figure 6 a block like structures separated by cracks are observed on the edge of the substrate which is attributed to the removal of solvent residuals during the drying process, and which may lead to hierarchical crack networks.[61] This phenomena was also observed for annealed films of bismuth oxide.[38] The volume reduction led to surface tensions and microcracking of the films. Additionally, in Figure 6b–e island growth can nicely be determined. Especially at the boarders of the substrate (Figure 6b) the islands occur as white spots in the SEM images, which is an indicator for charged structures and related to the height. The islands occur as result of the interaction between several molecules of compound **2.**

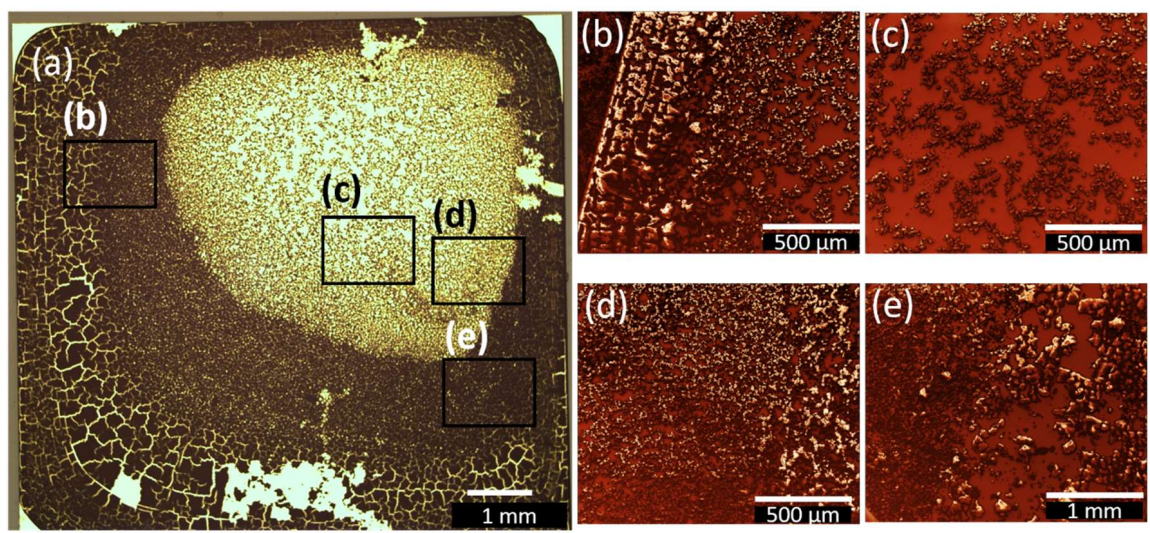

Figure 6. (a) Overview optical microscopy image for **2**-DrE-2 after the complete evaporation of ethanol. For better resolution SEM images with the same magnification were chosen in (b)–(e) at several spots of the sample, shown after applying image filter Gwyddion.net from Gwyddion software. The SEM images indicate island growth behavior.

In contrast to the growth behavior observed for **2**-DrE-2 from ethanol solution, the one obtained from acetonitrile is indicating diffusion limited aggregation in the central droplet region as presented in Figure 7. Fewer cracked structures occur on the borders than for **2**-DrE-2, most likely due to the slightly different evaporation rates of the two solvents. Nevertheless, the dominant growth mode in this sample is diffusion limited aggregation, since it is known to lead to the formation of characteristic dendritic, floral structures.[62] This occurs when diffusion on the surface is the primary mechanism of mass transport in the system and desorption-adsorption play a minor role.[62] The diffusion leads towards sticking of particles to the nucleation sites at the rim of the aggregates, where the floral-like structure starts growing. The floral-like agglomerations of **2** only occur during the drop and dip coating procedure.

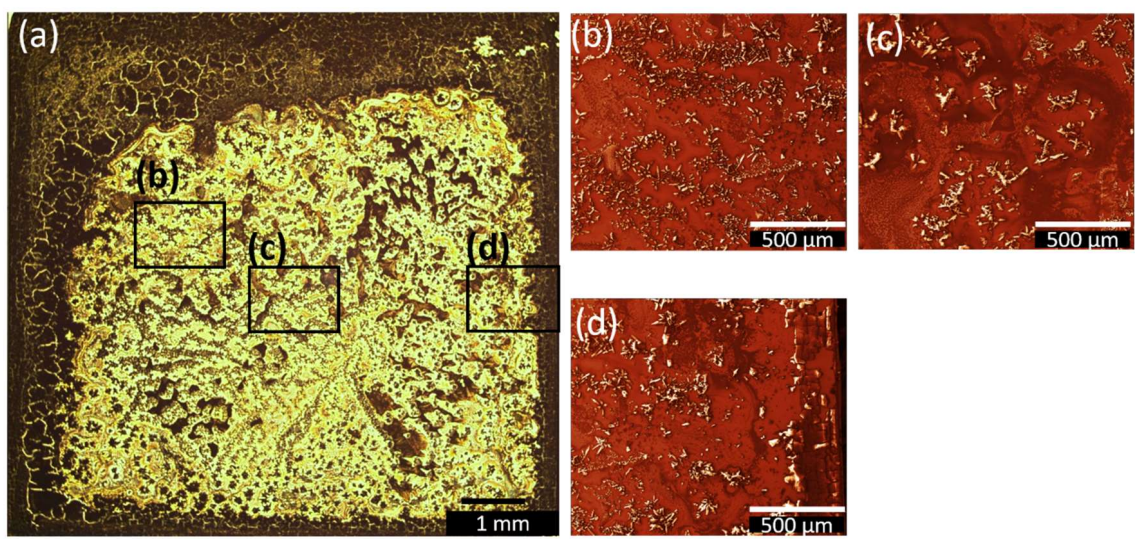

Figure 7. (a) Overview microscopy image for **2**-DrA-2 after complete evaporation of acetonitrile. For better resolution, SEM images with same magnification were chosen in (b)-(d) at several spots of the sample, shown after applying the image filter Gwyddion.net from Gwyddion software. The SEM images indicate areas with dendritic growth typical for diffusion limited aggregation.

**Spin coated and dip coated films**

In the following, a comparison between the films of the bismuth oxido cluster **2** deposited by spin and dip coating techniques is presented.

Optical microscopy images of spin coated and dip coated films from ethanol and acetonitrile solutions are shown in Figure 8 with a magnification of x1 as well as SEM images for a better resolution with a magnification of x2000. While the mechanisms discussed in ref.[58] for the formation of nanoparticle agglomerations can be well applied to describe the morphology of the films grown by drop coating (just for **2**-DrE and **2**-DrA), the images in Figure 8 indicate that films deposited by spin and dip coating are not subjected to severe concentration gradients during growth.

In the case of spin and dip coated samples the coffee stain effect was not observed. Compared to the case of drop coating, a tendency of agglomeration of compound **2** is observable, with the main difference that in the case of spin and dip coated samples homogeneously distributed agglomerations over the whole substrate are observed (for dip coating – over the whole dipped area). These, are highly reproducible. The homogeneous distribution of the agglomerations can be related to the faster spreading of the solution; hence the droplet size is reduced in the case of spin and dip coating. This, in turn, reduces the contribution of the radial flow towards the droplet edge and hence the radial concentration gradients.

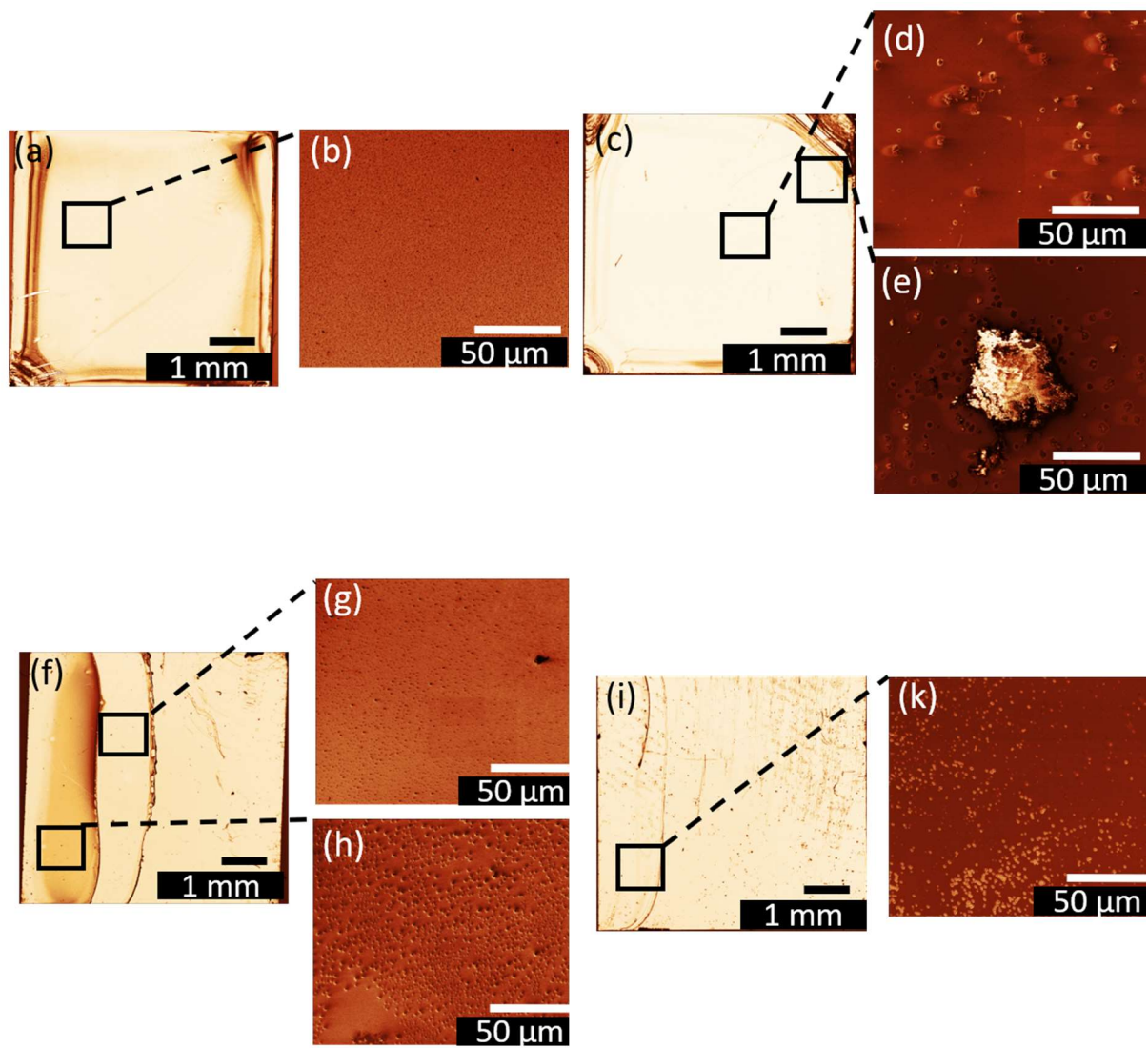

Figure 8. Light microscopy images with insets from SEM images of spin and dip coated films treated with the Gwyddion.net image filter from Gwyddion software for (a), (b) **2**-SpE, (c)-(e) **2**-SpA, (f)-(h) **2**-DiE and (i), (k) **2**-DiA.

The area covered by the single aggregates is larger in the case of acetonitrile than for ethanol solution (see Figure 8b and Figure 8d, respectively). The larger size of agglomerations grown from acetonitrile solution is consistent with the fact that the solubility in ethanol is more than twice as high as for acetonitrile. As the two solutions are equimolar, precipitation of **2** will start earlier from acetonitrile and the aggregates have more time to grow and coagulate. This effect is enhanced by the lower evaporation rate of acetonitrile compared to ethanol. Notably, the solubility experiments proved that the resulting agglomerates of **2** dissolved in ethanol are smaller due to the better solubility. Next to the island formation Ostwald ripening[62] was observed, especially for **2**-SpA (cf. Figure 8e). This phenomenon occurs in liquid environment and leads to bigger agglomerates separated by an empty zone from the smaller ones, which get dissolved during the growth process. Ostwald ripening is a thermodynamically driven process, which minimizes the island edge line and the high surface to volume ratio is decreased by the formation of big agglomerates. By comparing the particle area of the two samples obtained from ethanol and acetonitrile the difference between the two solvents used is quite obvious and the calculated particle density for acetonitrile samples is much lower than that of ethanol ones (cf.Table 2). The highest possible coverage of (59.95 ± 3.00) % was achieved for the **2**-SpE sample. Overall, full coverage of the substrate by **2** cannot be achieved with the chosen deposition parameters. The coverage extracted from all microscopy images shown in Figure 8 are presented in Table 2.

AFM measurements were performed to determine the height and the shape of the agglomerations. Micrometer-sized 3D aggregate shapes are obtained in all cases. The results are shown in Figure 9 and the associated maximum heights are summarized in Table 2. The floral structure grown during dip coating from ethanol solution (Figure 9c) shows flatter maximum height than the smaller islands formed (cf. Figure 9 a,b and d). The origin of the island height difference observed for the two deposition methods is not fully understood. The fractal boundary shape again suggests further growth on the gold surface by diffusion limited aggregation. In both cases indications for Ostwald ripening of larger particles at the expense of smaller ones are present. Depending on the solvent evaporation rate one or the other mechanism dominates, as discussed in more detail below.

For the spin coated samples shown in Figure 9Figure a and b islands well-defined terraces are formed. For the ethanol spin coated sample in Figure 9a combination of large and small islands can be observed. Interestingly, in the vicinity of large particles, a large area is depleted from islands; this dark red area of the image probably corresponds to pure Au. A "void" depletion region surrounding the cluster, which is typical for Ostwald ripening, can also be seen in the microscopy image of samples spin coated from acetonitrile in Figure 9b as light white shadow of the agglomerates. In some cases, those shadows, corresponding to the gold surface, are elliptically shaped with the long axis of the ellipse pointing in the same direction. It is well possible that this characteristic shape is a result of outward flow of the solution, fast evaporation time during the spinning process and the lower solubility of compound **2** in acetonitrile.

The dominant growth mode in the sample shown in Figure 9c is the diffusion limited aggregation.[62] The floral-like agglomerations of **2** only occurs during the drop (see above, different scales) and dip coating procedure. Therefore, it can be assumed that this process is related to the available time for the formation of this floral pattern. If the solvent is not fully evaporated during the retraction of the substrate, the molecules of **2** have sufficient time to agglomerate.

Summarizing, the deposition of bismuth oxido cluster **2**, from different solutions and using three different deposition methods can lead to the formation of agglomerations or islands of different shape and size. Considering the XRD results, these structures are crystalline.

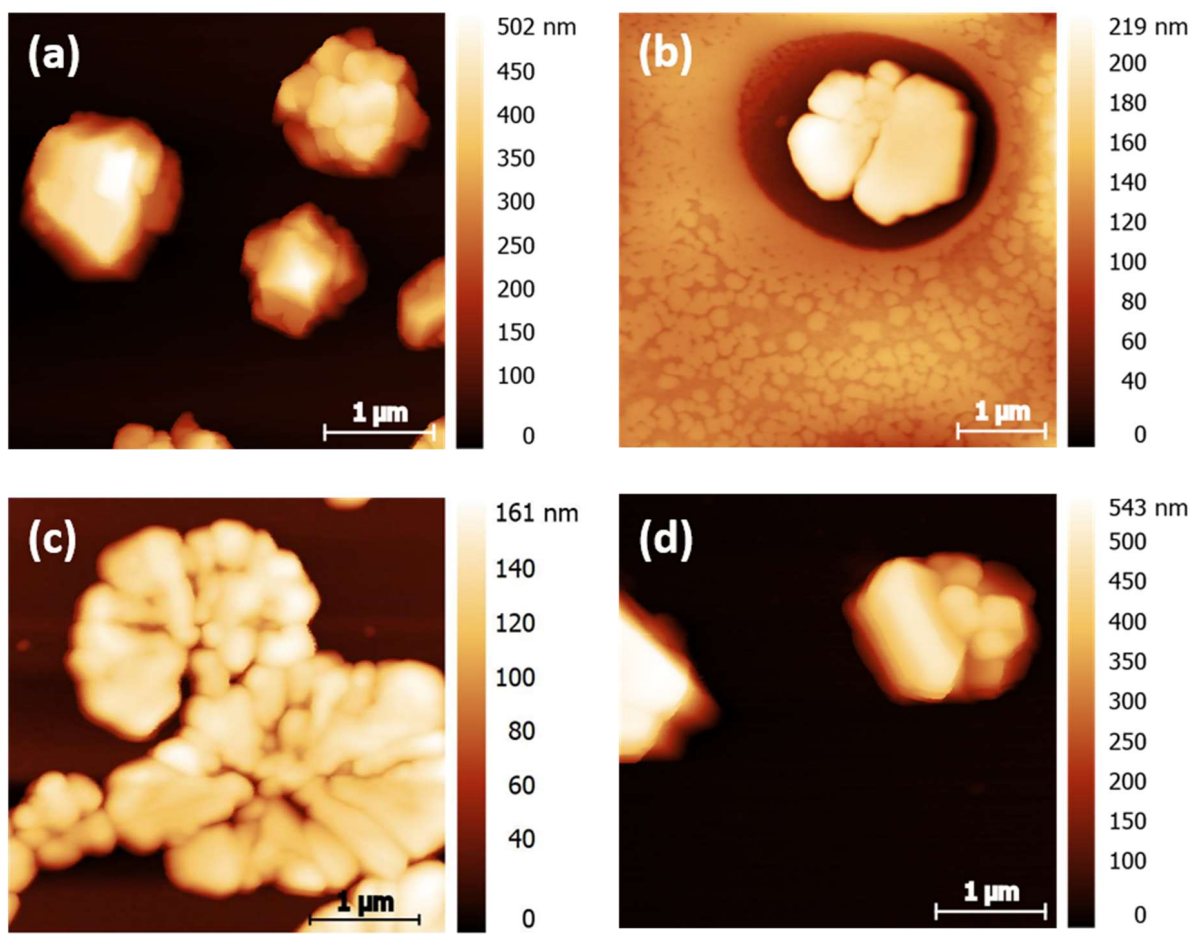

Figure 9. AFM images of (a) **2**-SpE, (b) **2**-SpA, (c) **2**-DiE, and (d) **2**-DiA films. In (a), (b), and (d) islands of sub-$\mu$m lateral size and heights are visible. Their density depends on the choice of the solvent and on the deposition method. In the case of dip coating in ethanol solution (c), a floral pattern is formed.

Table 2. Particle densities of the films extracted from the optical microscopy images shown in Figure 8 and maximum height for agglomerations extracted from the AFM images shown in Figure 9.

| sample | particle density / % (microscopy images) | maximum island height / nm (AFM) |
|---|---|---|
| **2**-SpE | (60.0 ± 3.0) | (502 ± 25) |
| **2**-SpA | (4.1 ± 0.4) | (219 ± 11) |
| **2**-DiE | (28.6 ± 1.4) | (161 ± 8) |
| **2**-DiA | (13.8 ± 0.2) | (543 ± 27) |

## Conclusions

In this work, the high yield conversion of the bismuth subnitrate [Bi$_{38}$O$_{45}$(NO$_3$)$_{20}$(dmso)$_{28}$](NO$_3$)$_4$·4dmso into the chiral bismuth oxido nanocluster [Bi$_{38}$O$_{45}$(Boc-Phe-O)$_{24}$(dmso)$_9$] (**2**) is demonstrated. The nanocluster **2** shows a hydrodynamic diameter of (1.4 – 1.6) nm (in CH$_3$CN) and (2.2 – 2.9) nm (in Ethanol) according to DLS. PXRD results are in line with a diameter of approximately 2 nm. Cluster **2** is soluble in a variety of organic solvents, which makes deposition methods from solution such as spin, dip and drop coating attractive. Using XRD, XPS and FTIR spectroscopy, it is demonstrated that the cluster core stays intact upon deposition on gold surfaces. Macroscopically, the films obtained by drop coating are neither homogeneous nor do they spread over a large area by using the same amount of solution compared to spin and dip coating. Optical microscopy as well as SEM images show that the growth behaviour in drop coating can be explained by the coffee ring effect. By the usage of more solution, cracked, very rough structures with a tendency to diffusion limited aggregation were observed. By using spin and dip coating methods, bismuth oxido cluster agglomerations with a lateral size on the micrometre scale were obtained. The lateral dimensions of the crystalline agglomerates are nearly monodisperse and the distance between the neighbouring micro-agglomerates is nearly constant. Since their size can be controlled by the choice of solvent and of the deposition method, compound **2** could be suitable for the growth of high-density bismuth oxido micro or nanostructures on prepatterned surfaces, where the prepatterning provides nucleation centres. Finally, this study provides a basis for further studies on wafer scale deposition of micro and nanostructures made of neutral metal oxido nanoclusters on metal surfaces, that are promising heterostructures for electronic or photonic applications.

**Acknowledgment**

We thank Jana Buschmann for CHNS analyses. A.S. and G.S. would like to acknowledge financial support from DFG project no. 282193534. We would like to thank Maria Almeida-Hofmann and Prof. Stefan Schulz for the possibility to use their light microscope.

**Supporting Information**.

- FTIR spectra of Boc–Phe sodium salt **1** (and [$Bi_{38}O_{45}$( Boc–Phe–O)$_{24}$(dmso)$_9$] **2** powder
- PXRD pattern of [$Bi_{38}O_{45}(NO_3)_{20}$(dmso)$_{28}$](NO$_3$)$_4$·4dmso (**A**, green) and [$Bi_{38}O_{45}$(Boc–Phe–O)$_{24}$(dmso)$_9$] powder (**2**)
- Exemplary PSD curves of [$Bi_{38}O_{45}(NO_3)_{20}$(dmso)$_{28}$](NO$_3$)$_4$·4dmso (**A, a**) and **2** in dmso (**b**), in acetonitrile (**c**) and in ethanol (**d**)
- Summary of the FTIR vibration modes of **2** in powder and films
- FTIR spectra for drop and spin coated films, **2** dissolved in ethanol (DrE and SpE) and acetonitrile (DrA and SpA)
- Comparison of XRD of compound **2** and drop coated films of **2** from ethanol (**2**-DrE) and acetonitrile (**2**-DrA) solution

**Corresponding Authors**

* salvan@physik.tu-chemnitz.de, michael.mehring@chemie.tu-chemnitz.de

**Author Contributions**

A.M. and R.T. contributed equally to this work. In particular, R.T. manufactured the **2** compound and provided the PXRD, NMR and DLS analysis. Furthermore R.T. provided necessary assignment of vibration modes for IR data. A.M. carried out the deposition experiments as well as light microscopy images and IR analysis. AFM was performed from

# Supporting Information

# Deposition of nanosized amino acid functionalized bismuth oxido clusters on gold surfaces

*Annika Morgenstern[1], Rico Thomas[2], Apoorva Sharma[1], Marcus Weber[2], Oleksandr Selyshchev[1], Ilya Milekhin[1], Doreen Dentel[4], Sibylle Gemming[1,3], Christoph Tegenkamp[4], Dietrich R.T. Zahn,[1,3], Michael Mehring[2,3], Georgeta Salvan[1,3]*

[1] Semiconductor Physics, Chemnitz University of Technology, 09107 Chemnitz, Germany

[2] Coordination Chemistry, Chemnitz University of Technology, 09107 Chemnitz, Germany

[3] Center of Materials, Architectures and Integration of Nanomembranes, Chemnitz University of Technology, 09126 Chemnitz, Germany

[4] Analysis of Solid Surfaces, Chemnitz University of Technology, 09107 Chemnitz, Germany

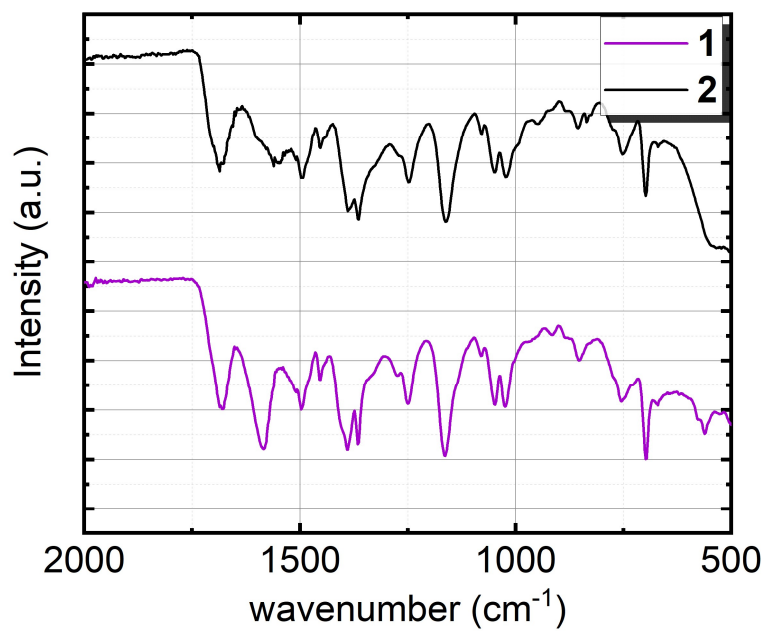

Figure S1. FTIR spectra of Boc–Phe sodium salt **1** (violet) and [$Bi_{38}O_{45}$( Boc–Phe–O)$_{24}$(dmso)$_9$] **2** powder (black).

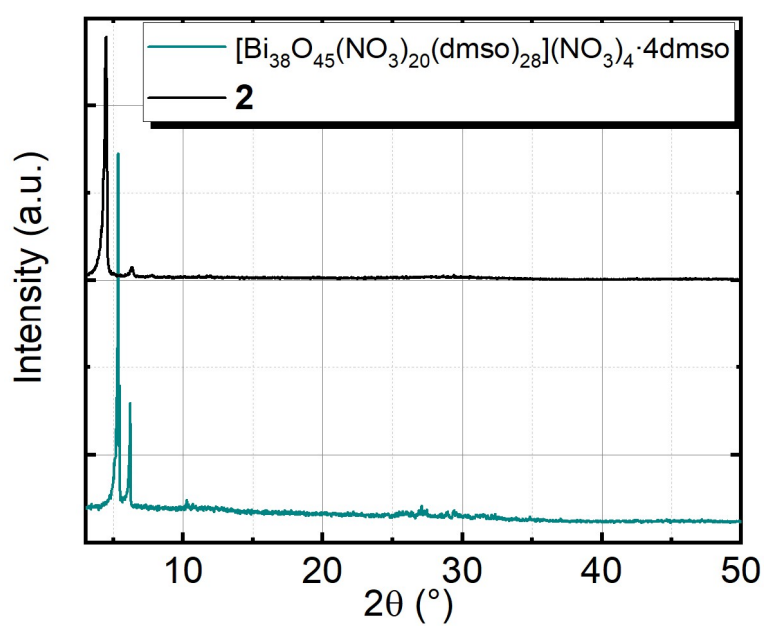

Figure S2. PXRD pattern of [$Bi_{38}O_{45}(NO_3)_{20}$(dmso)$_{28}$]($NO_3$)$_4$·4dmso (**A**, green) and [$Bi_{38}O_{45}$(Boc–Phe–O)$_{24}$(dmso)$_9$] powder (**2**, black).

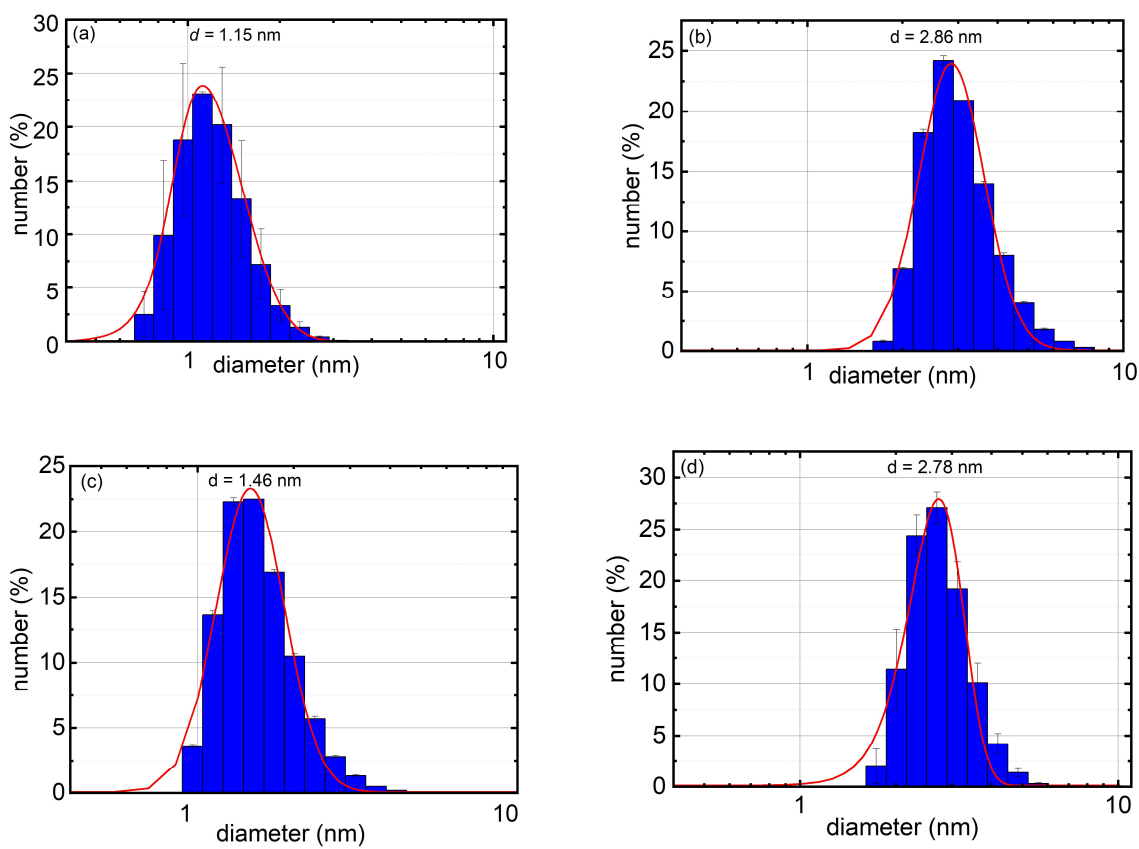

Figure S3. Exemplary PSD curves of $[Bi_{38}O_{45}(NO_3)_{20}(dmso)_{28}](NO_3)_4 \cdot 4dmso$ (**A**, **a**) and **2** in dmso (**b**), in acetonitrile (**c**) and in ethanol (**d**). Rare Date for PSD curve of **A** were used with permission from[63].

Table S1. FTIR vibration modes of **2** associated with FTIR spectrum in Figure 1 and Figure S4 (a) **2**, (b) **2**-DrA, (c) **2**-DrE, (d) **2**-DiA, (e) **2**-DiE, (f) **2**-SpA and (h) **2**-SpE (values given in cm$^{-1}$).

| Mode | (a) | (b) | (c) | (d) | (e) | (f) | (g) |
|---|---|---|---|---|---|---|---|
| $\nu_{Bi–O}$ | 582 | 576 | 583 | 576 | 570 | 575 | 577 |
| $\delta_{monosubst.\ Aromat}$ | 699, 753 | 699, 752 | 701, 754 | 700, 752 | 700, 753 | 699, 752 | 700, 752 |
| $\nu_{S=O}$ | 948 | 946 | 914 | 918 | 917 | 947 | 947 |
| $\delta_{C–H}(C=C–H)$ | 1018 | 1021 | 1026 | 1023 | 1027 | 1021 | 1022 |
| $\nu_{C–N} / \nu_{C–O}$ | 1162, 1047 | 1165, 1050 | 1170, 1053 | 1163, 1051 | 1166, 1051 | 1162, 1050 | 1164, 1051 |
| $\nu_{C–O–C}$ | 1246 | 1246 | 1251 | 1247 | 1249 | 1247 | 1247 |
| $\delta_{C–H}(CH_3)$ | 1363, 1386 | 1364, 1387 | 1368, 1393 | 1365, 1389 | 1366, 1391 | 1363, 1389 | 1365, 1389 |
| $\nu_{C=O}$ carboxylate | 1465, 1554 | 1495, 1553 | 1497, 1559 | 1496, 1557 | 1496, 1558 | 1495, 1554 | 1496, 1556 |
| $\nu_{C=O}$ amide | 1688 | 1685 | 1694 | 1684 | 1685 | 1689 | 1690 |
| $\nu_{as.C–H}$ | 2928, 2975 | 2929, 2976 | 2943, 2978 | 2916, 2976 | 2922, 2976 | 2929, 2977 | 2928, 2977 |

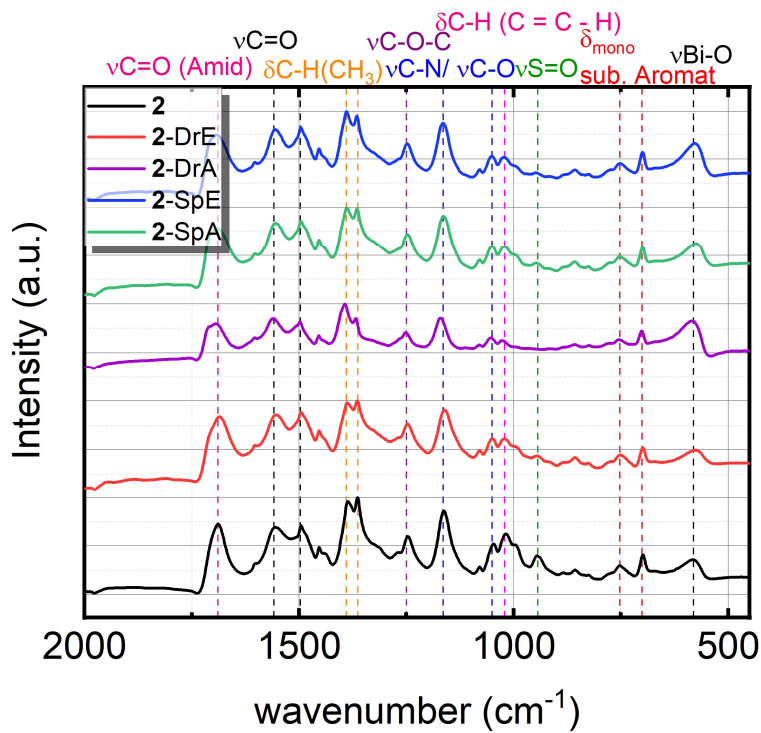

Figure S4. FTIR spectra for drop and spin coated films, **2** dissolved in ethanol (DrE and SpE) and acetonitrile (DrA and SpA).

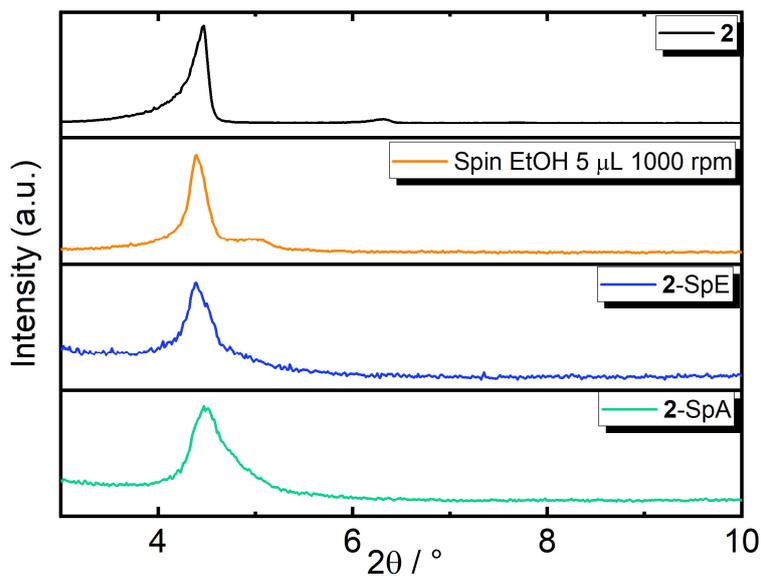

Figure S5. Comparison of XRD of compound **2** and drop coated films of **2** from ethanol (**2**-DrE) and acetonitrile (**2**-DrA) solution.

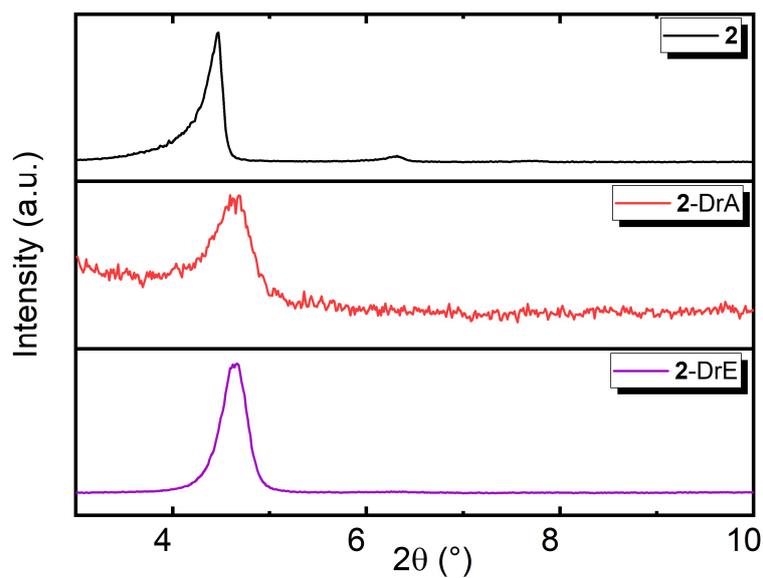

Figure S6. Comparison of XRD of compound **2**, spin coated samples of **2** from ethanol 5 µL and 1000 rpm, ethanol with 20 µL and 2000 rpm (**2**-SpE), and from acetonitrile 20 µL and 2000 rpm (**2**-SpA) solution.

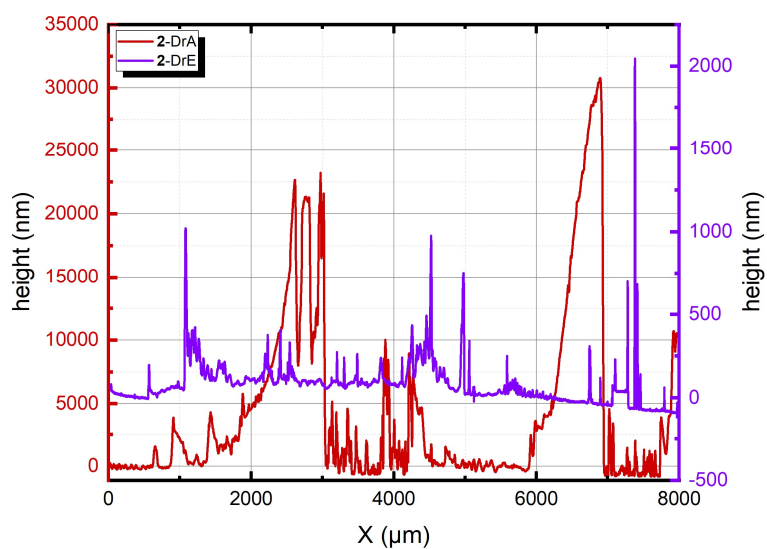

Figure S7. Profilometry of deposited films from bismuth oxido cluster **2** dissolved in ethanol (DrE) and in acetonitrile (DrA), respectively.